# Towards Guidelines for Preventing Critical Requirements Engineering Problems

Priscilla Mafra
Marcos Kalinowski
Computing Institute
Fluminense Federal University
Niterói, Brazil
pmafra@ic.uff.br
kalinowski@ic.uff.br

Daniel Méndez Fernández
Software and Systems Engineering
Technical University of Munich
Munich, Germany
daniel.mendez@tum.de

Michael Felderer
Institute of Computer Science
University of Innsbruck
Innsbruck, Austria
michael.felderer@uibk.ac.at

Stefan Wagner
Institute for Software Technology
University of Stuttgart
Stuttgart, Germany
stefan.wagner@informatik.uni-stuttgart.de

*Abstract*—[Context] Problems in Requirements Engineering (RE) can lead to serious consequences during the software development lifecycle. [Goal] The goal of this paper is to propose empirically-based guidelines that can be used by different types of organisations according to their size (small, medium or large) and process model (agile or plan-driven) to help them in preventing such problems. [Method] We analysed data from a survey on RE problems answered by 228 organisations in 10 different countries. [Results] We identified the most critical RE problems, their causes and mitigation actions, organizing this information by clusters of size and process model. Finally, we analysed the causes and mitigation actions of the critical problems of each cluster to get further insights into how to prevent them. [Conclusions] Based on our results, we suggest preliminary guidelines for preventing critical RE problems in response to context characteristics of the companies.

## I. INTRODUCTION

Requirements Engineering (RE) is highly volatile and inherently complex by nature [1]. Given this complexity, organisations may face several problems during the RE process, leading to severe implications, including project failure [2][3]. Furthermore, the cost for correcting RE-related problems increases throughout the software development life cycle [4], which reinforces the importance of introducing means to preventing problems right from the beginning of project. However, there are no specific guidelines that could be used by different types of organisations to help them in preventing critical RE problems.

In order to better understand the status quo on RE practice and its critical problems, a project named NaPiRE (Naming the Pain in Requirements Engineering) was launched in 2012 involving researchers from different countries [1]. The NaPiRE project comprises the design of a family of surveys on RE and its goal is to lay an empirical foundation about practical problems and needs of RE to allow directing future research in a problem-driven manner [1]. The last survey round (2014-2015) was answered by in total 228 organisations from 10 different countries around the globe, including organisations of different sizes and following different process models.

The goal of this paper is to use the data obtained from the NaPiRE initiative and to propose empirically-based guidelines that can be used by different types of organisations to support them in the prevention of critical RE problems. To this end, we identified the most critical RE problems, their causes and mitigation actions, grouping the organisations into clusters according to their size and process model. For each cluster, we selected the problems that were cited by the respondents as critical and leading to project failure. After selecting the most critical problems per cluster, we analysed their causes and mitigation actions to propose the guidelines to prevent them.

The remainder of this paper is organized as follows. Section 2 describes the background and related work. Section 3 describes the analysis of the NaPiRE data used to build the guidelines. Section 4 presents the resulting guidelines. Finally, Section 5 presents the concluding remarks and future work.

## II. BACKGROUND AND RELATED WORK

To lay the foundations for this paper, we discuss work related to survey research on RE problems and introduce the NaPiRE initiative and its previously published material.

### A. Survey Research on Requirements Engineering Problems

A well-known survey on causes for project failure is the Chaos Report of the Standish Group on cross-company root causes for project failures. While most of these causes are related to RE, the survey has serious design flaws and the validity of its results is questionable [5]. Moreover, unfortunately it does not directly support the investigation of RE problems in industry.

Some surveys have been focusing specifically on RE problems in industry. These surveys include the one conducted by Hall *et al.* [6] in 12 software organisations. Their findings, among others, suggest that most RE problems are mainly organisational rather than technical. Some researchers conducted country-specific investigations on RE problems. Solemon *et al.* [7] conducted a survey about RE problems in small organisations in Malaysia. Liu *et al.* [1] also conducted a study about RE problems in Chinese organisations.

However, each of those studies focused on specific aspects in RE and they mainly reported problems identified for their investigation scenario, without an in-depth discussion on their prevention. Additionally, the studies are completely independent and their results are isolated and not generalizable. To address these issues, the NaPiRE project[1] was launched [1].

*B. The NaPiRE Project*

The NaPiRE project started as a reaction to the lack of a general empirical basis for requirements engineering research [3]. This survey is currently being replicated in several countries around the globe[1]. The survey questionnaire contains 35 questions gathering the following type of data from the responding organisations: (a) general information, (b) RE status quo, (c) RE improvement status quo, (d) RE problems faced in practice, and (e) RE problem manifestation (e.g., causes, impact, and mitigation actions).

Some NaPiRE related research was already conducted. In [9] data from 74 Brazilian organisations was used to identify critical RE problems and their main causes. The most critical RE problems, according to those organisations, are related to communication and to incomplete/hidden or underspecified requirements. Furthermore, we provided probabilistic cause-effect diagrams [10][11] to organise knowledge on common causes of the five most critical identified RE problems.

In [12], an analysis of the similarities and the differences in the problems experienced in Brazil and Germany was made. In this paper it was possible to observe that the dominating factors are related to human interactions (e.g., based on the size and process model) rather than country. Again incomplete/hidden requirements and poor communication were among the most critical reported RE problems. Finally, in [13], we took a first step into RE problem prevention by further analysing the reported common causes and mitigation actions for the specific problem of incomplete/hidden requirements based on the Austrian and Brazilian data.

In this paper we significantly extend the effort done in [13] based on the findings in [12], by looking at the complete data set, identifying and analysing critical problems, causes and mitigation actions, independently of the country. Therefrom we aim at deriving guidelines for organisations to help them in preventing the problems.

## III. RESEARCH METHODOLOGY

In this section we describe the research method used to move from the NaPiRE data towards guidelines for preventing critical RE problems.

*A. Study Population*

In total, 354 organisations agreed to answer the NaPiRE survey. Out of these, 228 (63%) completed the survey by going through all of its questions and successfully reaching its end.

We decided to use only data of the 228 organisations that fully answered the questions. As we decided to build the guidelines focusing on the organisations by cluster of size (small, medium and large organisations) and process model (agile and plan-driven) some treatment had to be done on the data collected from the questionnaire. As in [12], we grouped the organisations by their number of employees. Organisations that have up to 50 employees were considered as small-sized; from 51 to 250 were considered medium-sized, and with more than 251 employees were considered large-sized. Out of the 228 organisations, 216 provided their number of employees.

[1] NaPiRE project web site http://www.re-survey.org/

Regarding the process models used, respondents answered a multiple-choice question with the following options: RUP, Scrum, V-Model XT, Waterfall, XP, and Other (informed textually) [3]. For our analysis, we grouped these process models into agile (Scrum and XP) and plan-driven (RUP, V-Model XT and Waterfall). Out of the 228 organisations, 196 selected one of the five predefined options as their process model. Out of these, 58 used mixed process models, i.e. reported their process fitting into options of both groups. We decided to exclude the mixed ones from our analysis to remove a potential confounding factor, as in these cases we had no information on the extent to which each process model is applied in the organisation.

*B. Data Analysis*

We first used the information concerning size and process model of the responding organisations and crossed them to explore potential RE problem patterns within each cluster, as also done in [3]. The result of this crossing is shown in Table I. 136 organisations reported to use agile or plan-driven process models and also informed their size (2 of the 138 agile and plan-driven organisations did not inform their size).

TABLE I. NUMBER OF ORGANISATIONS PER CLUSTER

| Process Model | Size | Total |
|---|---|---|
| Agile | Small | 30 |
| Agile | Medium | 22 |
| Agile | Large | 39 |
| Plan-driven | Small | 11 |
| Plan-driven | Medium | 4 |
| Plan-driven | Large | 20 |
| Total | | 136 |

Our goal on crossing those characteristics was to be more precise about the type of organisations, facilitating their use of the resulting guidelines, given that the types of problems faced by these organisations could be slightly different, as observed in [12]. After crossing the clusters, the next step was to analyse the most critical problems within each cluster.

The NaPiRE questionnaire presents respondents with a list of problems that, based on previously conducted studies [1], practitioners are meant to typically encounter in practice. Respondents were asked to report on the relevance of the presented problems for their project setting and then to list the five most critical ones. They were also asked to inform, for each of the critical problems, whether the problem leads to project failure or not.

Based on this information, we compiled, within each cluster, the total number of citations of each problem as a critical one and how often the problem was reported to lead to project failure. To keep the focus of the guidelines on the main problems, we decided that the guidelines should address the 5 most frequently cited problems within each cluster. Finally, unlike paper [3], we analysed the causes and mitigation actions reported by the respondents of each cluster for its most critical problems to build the guidelines.

## C. Top RE Problems, Causes, and Mitigation Actions

We compiled the 5 most critical RE problems within each cluster. The result is shown in Table II. It is noteworthy that medium and large-sized plan-driven organisations have 6 critical problems listed. This occurred because the same number respondents cited the last two listed problems.

TABLE II. MOST CRITICAL PROBLEMS WITHIN EACH CLUSTER

|  | Agile | Plan-driven |
|---|---|---|
| Small | • Incomplete and / or hidden requirements<br>• Communication flaws between us and the customer<br>• Underspecified requirements<br>• Communication flaws within the project team<br>• Time boxing | • Incomplete and / or hidden requirements<br>• Communication flaws within the project team<br>• Moving targets<br>• Time boxing / Not enough time in general<br>• Underspecified requirements |
| Medium | • Communication flaws between us and the customer<br>• Incomplete and / or hidden requirements<br>• Communication flaws within the project team<br>• Stakeholders with difficulties in separating requirements from previously known solution designs<br>• Weak access to customer needs and / or (internal) business information | • Communication flaws between us and the customer<br>• Incomplete and / or hidden requirements<br>• Moving targets<br>• Gold plating<br>• Underspecified requirements Weak access to customer needs and / or (internal) business information |
| Large | • Incomplete and / or hidden requirements<br>• Moving targets<br>• Communication flaws between us and the customer<br>• Time boxing<br>• Underspecified requirements | • Incomplete and / or hidden requirements<br>• Communication flaws between us and the customer<br>• Underspecified requirements<br>• Communication flaws within the project team<br>• Moving targets<br>• Stakeholders with difficulties in separating requirements from previously known solution designs |

For the purpose of this paper, extending the previous effort, we analysed the causes and mitigation actions for each of the 9 problems that appear as top 5 problems in the different clusters. When analysing the causes and mitigation actions for a problem in a given cluster, we only considered the answers on causes and mitigation actions provided by respondents within this same cluster. This decision was taken because we considered that the causes and mitigation actions could be considerably different for organisations in the different clusters.

Both, the causes and mitigation actions, were informed by the respondents in plain text as answers to open questions. Therefore, to analyse the answers given to the open question on causes and mitigation actions, we applied textual coding techniques as recommended by Grounded Theory [14], generating and peer-reviewing codes (representing key characteristics) for each of the open text answers. Due to space constraints, we present the codes for causes and mitigation actions reported for only one specific cluster. We chose agile and large-sized organisations, which was the cluster with most respondents. The codes for the causes and mitigation actions of the other clusters and their top 5 problems are available online[2]. Table III shows the causes and mitigation actions for agile and large-sized organisations. The coded causes and mitigation actions are shown together with the number of times each code was cited by the respondents.

TABLE III. CAUSES AND MITIGATION ACTIONS FOR THE TOP 5 PROBLEMS IN AGILE AND LARGE-SIZED ORGANISATIONS

|  | Causes | Mitigation Actions |
|---|---|---|
| Incomplete and / or hidden requirements | • Communication flaws between team and customer (1)<br>• Insufficient analysis at the beginning of the project (1)<br>• Insufficient resources (1)<br>• Lack of a well-defined RE process (1)<br>• Missing IT project experience at customer side (1)<br>• Missing knowledge about development framework (1)<br>• Missing of a global view of the system (1)<br>• Missing requirements specification template (1)<br>• Poor requirements elicitation techniques (1)<br>• Requirements remain too abstract (1)<br>• Unavailability of requirements engineer (1)<br>• Unclear business needs (1)<br>• Unclear roles and responsibilities at customer side (2) | • Create a Definition of Readiness (1)<br>• Creation of requirements specification template (1)<br>• Greater customer commitment (1)<br>• Implementation of change management process (1)<br>• Integrate Testing and RE (1)<br>• Introduction and use of check lists for monitoring requirements along their life cycles (1)<br>• Introduction of an early feedback loop with customer (1)<br>• More analysis before commitment (1)<br>• Planning and execution of regular communication events/ meetings (1)<br>• Planning and execution of training (in order to improve skill and performance) (1) |
| Moving targets | • Changing business needs (3)<br>• Complexity of domain (1)<br>• Customer does not know what he wants (2)<br>• Insufficient information (1)<br>• Insufficient resources (1)<br>• Missing completeness check of requirements (1)<br>• Missing concentration on business needs (1)<br>• Poor project management (1)<br>• Poor requirements elicitation techniques (1)<br>• Unclear business needs (1)<br>• Volatile industry segment that leads to changes (1)<br>• Weak management at customer side (1) | • Better support from project management (1)<br>• Customer orientation (1)<br>• Explain impact of changes to customers. (1)<br>• Have an agile project management (1)<br>• Implementation of change management process (1)<br>• Increase awareness to focus on business processes (1)<br>• Introduction and use of check lists for monitoring requirements along their life cycles (1)<br>• Planning and execution of training (in order to improve skill and performance) (1)<br>• Work with open scope (1) |
| Communication flaws with the customer | • Communication flaws between team and customer (1)<br>• Complexity of domain (1)<br>• Conflicting stakeholder viewpoints (1)<br>• Insufficient agility (1)<br>• Insufficient resources (1)<br>• Language barriers (2)<br>• Missing customer involvement (1)<br>• Missing direct communication to customer (2)<br>• Subjective interpretations (1)<br>• Too high team distribution (1) | • Definition of a common structure to describe and explain requirements (1)<br>• Greater customer commitment (4)<br>• Introduction of an agile methodology (1)<br>• Introduction of an early feedback loop with customer (3)<br>• Planning and execution of regular communication events/ meetings (2)<br>• Planning and execution of training (in order to improve skill and performance) (1)<br>• Prototyping (1)<br>• Use of mock-ups (1) |
| Time boxing | • Complexity of RE (1)<br>• High workload (1)<br>• Insufficient resources (1)<br>• Lack of time (2)<br>• Policy restrictions (1)<br>• Poor project management (1)<br>• Solution orientation (1)<br>• Strict time schedule by customer (3)<br>• Unfeasible goals (1) | • Have an agile project management (1)<br>• Introduction and use of check lists for monitoring requirements along their life cycles (1)<br>• Introduction of an agile methodology (1)<br>• More communication with customers (1)<br>• Prioritization of activities / goals (1)<br>• Smaller project with defined time and goal (1) |

---

[2] Causes, mitigation actions and guidelines for the top 5 problems within each cluster: http://www.ic.uff.br/~kalinowski/seaa2016

| | Causes | Mitigation Actions |
|---|---|---|
| **Underspecified requirements** | • Customer does not know what he wants (1)<br>• Insufficient information (1)<br>• Insufficient resources (1)<br>• Language barriers (1)<br>• Missing knowledge about development framework (1)<br>• Missing requirements specification template (1)<br>• Requirements remain too abstract (1)<br>• Unavailability of requirements engineer (1) | • Creation of requirements specification template (3)<br>• Definition of a common structure to describe and explain requirements (1) |

The next step was to analyse the causes and mitigation actions to gather further insights into the prevention of the RE problems, assembling the results into candidate guidelines. The outcome of this analysis will be presented in the next section.

## IV. GUIDELINES FOR PREVENTING CRITICAL RE PROBLEMS

The candidate guidelines within each cluster were obtained through analysis and discussions conducted by the team involved in the paper. These analysis and discussions were based on the organised information on causes and mitigation actions for RE problems within each cluster (cf. Table III) and focused on how to prevent the problems.

During the discussions, when facing large lists of causes we organised them into Ishikawa diagrams [15] using the cause categories suggested in guidelines for defect causal analysis [16]. An exemplary diagram for the causes cited in agile and large organisations for the RE problem *Incomplete and/or hidden requirements* is shown in Figure 1. The suggested mitigation actions aligned with the proposed process model, on the other hand, were used as a starting point for the discussion on how to address the causes and prevent the problem.

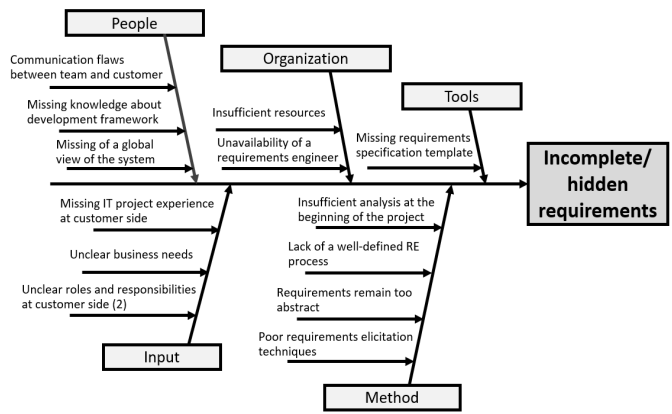

Fig. 1. Ishikawa diagram with causes of Incomplete and/or hidden requirements as reported by agile and large organisations.

For the problem depicted in Figure 1., we decided to include the advice in the guidelines to *conduct regular meetings with the customer, to introduce an early feedback loop with the customer* and to *provide training* (if needed) on the development framework and on the overall system scope. This shall address the causes in the *People* category, which were mainly related to communication flaws and missing knowledge on the development framework and of the overall system.

For causes of category *Organization, we inferred* the advice to explicitly *allocate a requirements engineer to the project*. For causes of category *Tools*, the suggestion is to *use a requirements specification template*. We are aware that it may be difficult to address the causes in the *Input* category, as addressing these causes is related to introducing changes at the customer side. Therefore, no specific advice was included for those causes.

Finally, as for the causes in the *Method* category, the causes *insufficient analysis at the beginning of the project* and *requirements remain to abstract* are addressed by an advice that follows one of the suggested mitigation actions, to *create a DoR (Definition of Readiness)*. Such DoR is commonly used in agile projects to avoid the beginning of work on features that do not comply with clearly defined completion criteria, which usually translates into costly rework. Additionally, also following mitigation actions listed by the respondents, we included advice to *spend more effort in requirements analysis and validation*. For the cause *poor requirements elicitation techniques* and *lack of a well-defined RE process,* besides the advice already provided, we included the suggestion to *implement a change management process*. Nevertheless, it is noteworthy to mention that explicitly defining a rigorous RE process is not well aligned to the chosen agile development paradigm.

The preliminary guidelines compiled for the critical RE problems faced by agile and large organisations can be seen in Table IV. The tables with the guidelines for the other clusters are available online[2].

TABLE IV. GUIDELINES FOR PREVENTING RE PROBLEMS IN AGILE AND LARGE-SIZED ORGANISATIONS

| Critical Problems | Guidelines (Advice) for Agile and Large-sized Organisations |
|---|---|
| Incomplete and / or hidden requirements | • Allocate a requirements engineer to the project (with high degree of experience and expertise)<br>• Conduct regular meetings with the customer<br>• Create a DoR (Definition of Readiness)<br>• Create a requirements specification template<br>• Implement a change management process<br>• Introduce an early feedback loop with customer<br>• Provide training (if needed) to improve team skills on the development framework<br>• Provide training (if needed) on the business domain and overall system scope<br>• Spend more effort in elicitation and analysis<br>• Spend more effort in requirements validation |
| Moving targets | • Allocate a project manager with high degree of experience and expertise<br>• Allocate a requirements engineer to the project (with high degree of experience and expertise)<br>• Conduct regular meetings with the customer<br>• Explain customers the impact of changes<br>• Focus on business needs<br>• Implement a change management process<br>• Provide training (if needed) on the business domain and overall system scope<br>• Spend more effort in elicitation and analysis<br>• Use iterative development with several increments (or sprints) |
| Communication flaws between us and the customer | • Allocate a project manager with high degree of experience and expertise (in particular, related to risk management)<br>• Allocate a requirements engineer to the project (with high degree of experience and expertise)<br>• Conduct regular meetings with the customer<br>• Explain customers the importance of their contribution<br>• Introduce an early feedback loop with customer<br>• Raise the level of abstraction with the customer |

| Critical Problems | Guidelines (Advice) for Agile and Large-sized Organisations |
|---|---|
| | • Spend more effort in elicitation and analysis Spend more effort in requirements specification<br>• Use prototyping |
| **Time boxing / Not enough time in general** | • Allocate a project manager with high degree of experience and expertise (in particular, related to risk management)<br>• Allocate the team members to work no more than 40 hours a week to maintain the productivity<br>• Conduct daily stand-up meetings with the project team<br>• Conduct regular meetings with the customer<br>• Implement a change management process<br>• Manage task priority<br>• Provide training (if needed) to improve team skills |
| **Underspecified requirements that are too abstract and allow for various interpretations** | • Allocate a requirements engineer to the project (with high degree of experience and expertise)<br>• Allocate a project manager with high degree of experience and expertise<br>• Conduct regular meetings with the customer<br>• Create a requirements specification template<br>• Use prototyping<br>• Provide training (if needed) to improve team skills<br>• Spend more effort in requirements specification<br>• Spend more effort in requirements validation |

The presented guidelines require additional review by experts from industry to evaluate their adequacy to address specific RE problems in specific contexts. Also, the guidelines need further refinement by observing their application in controlled experiments and industrial case studies so that they fit the particularities of industrial context and the practices used therein.

Finally, there are also threats to validity of the presented guidelines, which we mitigated by specific measures. The major threat to validity arises from the actual coding process of causes and mitigation actions to derive the guidelines. Coding is essentially a creative task with subjective facets of coders like experience, expertise and expectations, which we controlled by peer-reviewing the coding process. Another threat to validity comprises the misunderstanding/bias of the survey participants and the fact that, on the survey, the size of the organisations does not represent the size of the projects. The threat concerning to the size of the projects will be adjusted for the next execution of the survey. The underlying survey itself went through several validation cycles to reduce threats to validity [1]: the survey was built on the basis of a theory induced from available studies, internally and externally reviewed a few times, and piloted in an industrial context.

## V. CONCLUDING REMARKS

In this paper, we used data from the NaPiRE initiative to analyse the most critical RE problems for several types of organisations, as well as their common causes and mitigation actions suggested by the respondents. We organised the information by clusters of size and process model and, based on this data, suggested preliminary guidelines for preventing critical RE problems within each cluster.

The rationality and the process for the production of these resulting guidelines, allowed us to conclude that they can be useful for the organisations to avoid critical RE problems in practice due to the fact that the guidelines were produced based on the source of each problem, analysing their causes and the mitigation actions suggested by the respondents. However, despite the potential practical impact, we are aware that the guidelines need further evaluation. Future work therefore includes additional expert reviews of the proposed guidelines and empirical evaluations through experiments and case studies to increase their accuracy to which we cordially invite researchers and practitioners to strengthen our body of knowledge in preventing common problems in RE.

ACKNOWLEDGMENT

The authors would like to thank the NaPiRE community for their support. Thanks also to the Brazilian Research Council (CNPq, process number 460627/2014-7) for financial support.